\documentstyle[12pt,aasms4]{article}

\def\km/s{km~s$^{-1}$}
\def\um{${\mu}$m}

\def\Lo{L$_\odot$}
\def\Vinf{\hbox{$V_\infty$}}
\def\HeI{He\,{\sc i}}
\def\HeII{He\,{\sc ii}}

\def\NaI{Na\,{\sc i}}
\def\MgII{Mg\,{\sc ii}}
\def\FeII{Fe\,{\sc ii}}

\def\SiII{Si\,{\sc ii}}
\def\Mdot{\.{M}}
\def\FMM362{FMM\#362}

\slugcomment{to be published in The Astrophysical Journal Letters}
\lefthead{Geballe et al.}
\righthead{A Second LBV in the Quintuplet Cluster}
\begin{document}
\title{A Second Luminous Blue Variable \\ in the Quintuplet
Cluster}

\author{T. R. Geballe}
\affil{Gemini Observatory, 670 N. A'ohoku Pl., Hilo, HI 96720}
\authoremail{tgeballe@gemini.edu}

\author{F. Najarro}
\affil{Instituto de Estructura de la Materia, CSIC, Serrano 121, 28006
Madrid, Spain}

\and
\author{D. F. Figer}
\affil{Space Telescope Science Institute, 3700 San Martin Drive,
Baltimore, MD 21218}

\begin{abstract} 

H and K band moderate resolution and 4~\um\ high resolution spectra have
been obtained for \FMM362, a bright star in the Quintuplet Cluster near
the Galactic Center. The spectral features in these bands closely match
those of the Pistol Star, a luminous blue variable and one of the
most luminous stars known.  The new spectra and previously-obtained
photometry imply a very high luminosity for \FMM362, L~$\geq~10^6$~\Lo,
and a temperature of 10,000 -- 13,000~K. Based on its luminosity,
temperature, photometric variability, and similarities to the Pistol Star,
we conclude that \FMM362\ is a luminous blue variable.

\end {abstract}


\section {Introduction}
	
The Quintuplet Cluster (AFGL~2004), roughly 30~pc in projection from the
nucleus of the Galaxy, contains a number of massive and luminous stars
which are not detected at visible wavelengths due to heavy extinction by
dust along the line of sight. Some of the brightest of these stars are
enshrouded by circumstellar dust and have featureless infrared spectra,
apart from interstellar absorption bands (Okuda et al. 1990). Others are
not enshrouded and show line absorption and emission from their
photospheres and winds. Cotera et al. (1996) and Figer et al. (1998,
hereafter F98) have demonstrated that one such object, originally reported
as object No. 25 by Nagata et al. (1993), first singled out by Moneti,
Glass \& Moorwood (1994), and now known as the Pistol Star, has a
luminosity of $\sim~10^7$~\Lo, making it one of the most luminous stars
known. Figer, McLean and Morris (1995) suggested that this star is a
luminous blue variable (LBV), a hypothesis supported by its position in
the HR diagram (F98), photometric variability (Glass et al 1999; Figer,
McLean, \& Morris 1999a, hereafter FMM99), and circumstellar ejecta (Figer
et al. 1999b).

FMM99 have recently identified a second candidate LBV in the Quintuplet
Cluster, their source 362, hereafter \FMM362. The identification was based
on the star's infrared brightness, photometric variability (confirmed by
Glass et al. 1999), and a low-resolution spectrum obtained by Figer (1997,
unpublished).  Photometry by FMM99 and by Glass et al. suggest that at
maximum \FMM362\ is nearly as bright as the Pistol Star.

\section {Observations and Data Reduction}
 
We have obtained spectra of \FMM362\ in the H and K bands and near 4~\um\
at UKIRT with the facility 1-5~\um\ grating spectrometer, CGS4, which was
configured with a 256x256 InSb array and a 0.6'' (one-pixel) wide slit. An
observing log is provided in Table~1. The nearby featureless Quintuplet
source GCS~3-2 (Nagata et al 1990, Okuda et al. 1990; also listed as
source 2 by Glass et al 1990; as source 24 in Nagata et al 1993 and source
VR 5-2 in Moneti, Glass \& Moorwood 1994) was used as a comparison star,
although it is a suspected variable (Glass et al. 1999).

The H and K spectra were wavelength-calibrated with the aid of arc lamp
spectra.  The 4~\um\ spectrum was wavelength-calibrated by comparison to
the spectrum of the planetary nebula NGC~6572 ($V_{hel}$ = $-9$~\km/s).
Flux calibration in the H and K bands assumed that the dereddened spectrum
of GCS~3-2 is that of a 889~K blackbody (Okuda et al. 1990) with K~=~6.28
(Glass et al. 1999). Because of the uncertainty in this approximation, the
line fluxes far from 2.2~\um\ and the overall spectral shape are probably
not accurate. From our data we derive K=7.5 for \FMM362. The brightness is
consistent with previous photometry.  We believe that our relative
spectrophotometry is accurate to $\pm$20\%.
 
\section{Results and Initial Analysis}

The low resolution K-band spectrum of \FMM362\ from 1999 is shown in
Fig.~1 and the higher resolution spectra are shown in Figs. 2 -- 4.
Measured parameters of detected lines are given in Table~2. We note a
modest, but significant weakening of those spectral lines (2.10--2.18~\um)
observed on both dates. In spectral intervals where both the Pistol Star
and \FMM362\ have been measured, their spectra are quite similar. In
addition to lines of hydrogen, the same permitted lines of sodium,
magnesium, and iron are in emission in both stars and the He I lines,
where clearly detected, are in absorption. The principal difference
between the spectra is the lower equivalent widths of lines (in particular
those of hydrogen) in \FMM362. A possible additional difference between
the two stars is that forbidden lines are clearly seen only in the Pistol
Star. However, these are weak and, considering the smaller equivalent
widths of the lines in \FMM362, non-detections there are probably not
surprising.

The 4~\um\ spectrum (Fig.~2) is dominated by a strong hydrogen Br~$\alpha$
line at 4.05~\um, which because of its strength and equivalent width
appears more suitable than other lines for providing accurate velocity
information. The \HeI\ 5-4 triplet line, shifted $-240$~\km/s\ relative to
Br~$\alpha$, is weak but clearly present. Although the core of the
Br~$\alpha$ line is symmetric, even after allowing for the \HeI\ line
there is considerably more emission at high negative velocities than at
high positive velocities. This is caused by continuum opacity (Najarro et
al 1998a), which weakens the redshifted emission wing, an effect also
evident in the Pistol Star (Figer et al. 1998). We estimate that in
\FMM362\ the wings extend roughly to -250~\km/s and +125~\km/s\ from the
peak, somewhat further than those of the Pistol Star. The velocity of peak
emission is +121~$\pm$~15~\km/s\ (LSR).  This is very close to the
velocity of 130~\km/s\ determined for the Pistol Star (F98) and other
members of the cluster (Figer 1995), clearly establishing \FMM362\ as a
cluster member.

A number of \SiII, \MgII\ and \FeII\ lines are prominent in Figs. 3-4. The
\SiII\ 5s$^{2}$S$_{1/2}$-5p$^{2}$P$_{3/2}$~1.691~\um\ and
5s$^{2}$S$_{1/2}$-5p$^{2}$P$_{1/2}$~1.698~\um\ doublet is a powerful
diagnostic tool, as it appears in emission for only a very narrow range of
stellar temperatures and wind density structures, indicating the presence
of amplified NLTE effects (Najarro et al., in preparation). Several of the
\MgII\ lines observed in the H and K-bands have the 5p$^2$P level in
common. Those with it as the upper level (the 2.13/14~\um\ and
2.40/41~\um\ doublets) are much stronger than those with it as a lower
level (in the H band), revealing that pumping through the resonance
3s$^2$S-np$^2$P lines must be a significant populator of the np$^2$P
levels. Two types of lines are found for \FeII: the so-called
semi-forbidden lines (denoted in Table~2 and in the figures by single
left-hand brackets) such as \FeII\ z$^{4}$F$_{9/2}$-c$^{4}$F$_{9/2}$
1.688~\um\ and \FeII\ z$^{4}$F$_{3/2}$-c$^{4}$F$_{3/2}$~2.089~\um\ with
very weak oscillator strengths (gf$\sim$10$^{-5}$) which form in the outer
stellar wind; and permitted (gf$\sim$1) lines connecting higher lying
levels, such as the e$^{6}$G-5p$^{6}$F lines near 1.733~\um, which form
much closer to the atmosphere.

Except for Br~$\alpha$ (5--4), the rest of the observed Brackett series
lines (11--4, 10--4, and 7--4) show P Cygni profiles with the emission
strengthening and absorption weakening with decreasing series number. This
is expected in a dense wind where line emission increasingly overwhelms
the absorption profiles for lower series (higher oscillator strength)
lines, since these form further away from the photosphere. The same trend
is seen in the Humphreys series (14--6) hydrogen line at 4.02~\um. Because
of non-negligible continuum opacity effects at 4~\um, Br~$\alpha$ can only
provide a lower limit to \Vinf. The
\FeII~z$^{4}$F$_{9/2}$-c$^{4}$F$_{9/2}$ 1.688~\um\ line profile is much
less influenced by opacity effects. From it we estimate \Vinf\ to be
$\approx$160~\km/s.

Finally, it is noteworthy that the 1.700~\um\ and 2.112~\um\ lines of
\HeI\ appear weakly in absorption, while the \HeI~2.06~\um\ line is not
convincingly detected and the \HeI (5--4) emission components around
Br~$\alpha$ are very weak. This behavior, together with the complete
absence of \HeII\ lines, is a strong indicator of a low temperature (and
low ionization state).

\section{Discussion}

\subsection{\FMM362\ as an LBV} 

A rough estimate of basic stellar parameters of \FMM362\ can be made by
comparison of its spectrum with that of the Pistol Star, for which values
have been derived (F98). In particular, the presence of the \HeI\ lines in
absorption as well as observed ratio of the Fe~II 2.089~\um\ and
\MgII~2.14~\um\ lines tightly constrain the parameters of \FMM362\ as they
did for the Pistol Star. For the Pistol Star two families of models with
T$_{eff}$ = 14,000~K, L = 10$^{6.6}$\Lo\ and T$_{eff}$ = 21,000~K, L =
10$^{7.2}$~\Lo\ fit the R~$\approx$~1,000 infrared spectra. New higher
resolution spectra of the Pistol Star (Najarro et al. 1998b), when
analyzed with the model atmospheres of Hillier and Miller (1998), favor
the lower temperature solution. Given the similarities in their spectra,
the effective temperature of \FMM362\ probably also is low; additional
support for this is given in section 4.3.  Monitoring has shown that the
stars are nearly the same average brightness at K, whereas in 1996 the
Pistol Star was 0.5 mag brighter than \FMM362\ at J band (Figer, McLean
and Morris 1996). Assuming that the two stars have the same temperature
and that in \FMM362\ the infrared bound-free and free-free excess in the
continuum is negligible (as is the case for the Pistol Star below 3~\um),
their luminosity ratio is given by the extinction-corrected flux ratio.
From the close proximity of the two stars it is reasonable to assume that
the extinctions are the same, indicating that \FMM362\ is at least half as
luminous as the Pistol Star.

\subsection {Two LBVs in the Quintuplet Cluster?}

As there are only about a half dozen LBVs known in the Galaxy (Nota et al.
1995), one must question the identification of two LBVs in a single
cluster. However, the Quintuplet is one of the most massive young clusters
in the Galaxy, containing over 150 O-stars at birth (FMM99) and LBVs are
thought to be evolved O-stars (see Langer et al. 1994 for one proposed
evolutionary sequence).  The cluster age, 4~Myr (FMM99), is that when
O-stars should be evolving through the LBV stage. The number of cluster
LBV stars at any time during this stage is roughly N$_{LBV}$ =
($\tau_{LBV}$lifetime)/($\tau_{LBV-p}$/N$_{O-stars}$) = 25,000 yr/(6
Myr/150 O-stars)  = 2/3, where $\tau_{LBV-p}$ is the production timescale.
There are many uncertainties in this estimate: e.g., the assumptions that
all O-stars become LBVs; that the LBV lifetime is roughly equal to the
ratio of known galactic LBVs to known galactic O-stars times a typical
O-star lifetime; and that the O-stars become LBVs at a constant rate.
Nevertheless, we conclude that it is not unreasonable to find two LBVs in
this cluster.

\subsection{Qualitative analysis}

The equivalent widths of the emission lines in \FMM362\ are lower than in
the Pistol Star, not only for hydrogen, but also for \SiII, \MgII, \NaI, and
\FeII. This suggests that the stellar wind of \FMM362\ is less dense than
that of the Pistol Star. Our inference that the bound-free and free-free
contributions to the continuum are insignificant below 3~\um\ lead us to
conclude that the equivalent widths of the emission lines are proportional
to the wind density (see Simon et al 1983 or Najarro 1995) and hence that
the value of \Mdot\Vinf/R$^{2}$ for \FMM362\ is roughly a factor of two less
than for the Pistol Star. From luminosity considerations we estimate
(R*$_{Pistol}$/R*$_{362})^{2}$~$\approx$~1.5.  Taking into account the
scaling equations for \Mdot\ and R (Najarro et al 1997) and that
\Vinf$_{362}$~$\approx$~160~\km/s (from the \FeII\ 1.688~\um\ line) we
conclude that \Mdot$_{Pistol}$~$\approx$~1.5~\Mdot$_{362}$.

Important constraints on the stellar temperature (ionization) are set by
the weakness of the \HeI~2.06~\um\ line and especially by the weakness of
the \HeI\ (5-4) components near 4.05~\um. The observed ratio of the H and
\HeI\ component exceeds by a large factor the expected value even for
cosmic \HeI\ value, if there were any significant amount of \HeII\ in the
wind. This indicates that \HeII\ must recombine to \HeI\ very close to the
photosphere, implying an upper limit of around 13,000~K for the
temperature of the object. A lower limit on the effective temperature is
set by the non-detection of the
3s$^2$3p$^2$S$_{1/2}$-3s$^2$4p$^2$P$_{3/2}$~2.180~\um\ and
3s$^2$3p$^2$S$_{1/2}$-3s$^2$4p$^2$P$_{1/2}$~2.209~\um\ intercombination
lines of \SiII\ in absorption (note that the latter would be contaminated
by \NaI\ emission), since these lines are expected to be in absorption if
the temperature is below 10,000~K.

From the observed weak \HeI\ lines one might easily conclude that helium
is not enhanced at all. However, preliminary analysis (Najarro et al, in
preparation) shows that an enhancement as large as He/H$\sim$1 can be
completely masked. Only the 2.06~\um\ line is strengthened by increasing
He/H to this value, but the strength of this line also depends strongly on
blanketing. The spectrum of \FMM362 at 2.06~\um\ has been observed
at low resolution only.  Therefore, although we suspect that the value of
He/H is not far from normal, we cannot rule out a much higher value.

Finally, we consider the abundances of Mg, Si, and Fe. The striking
similarity of the \MgII\ lines in \FMM362\ to those in the Pistol Star
support a higher than solar Mg abundance for \FMM362\ (Najarro et al
1998b). The case of the \SiII\ lines is different. Although in principle a
high Si abundance is needed to obtain the H band lines in emission, the
behavior of these lines is controlled to first order by the wind density
structure and to second order by the effective temperature. Different
velocity fields and transition zones between photosphere and wind can
easily mask a factor of five change in Si abundance (producing similar
\SiII~1.7~\um\ emission line strengths and profiles), even if the stars
have the same effective temperature, radius, and mass-loss rate.

To derive the iron abundance two sets of \FeII\ lines can be used. The
z$^{4}$F-c$^{4}$F lines are formed in the outer regions of the wind where
the levels suffer from severe departures from LTE. Line strengths are
extremely sensitive to the ionization structure of Fe in the outer wind.
Initial tests (Najarro et al 1998b) showed that a high iron abundance is
required to reproduce these semi-forbidden lines. However, if charge
exchange reactions Fe$^{++}$~+~H~$\longleftrightarrow$~ Fe$^{+}$~+~H$^+$
are included (see Oliva et al 1989 and Hillier 1998), the ionization
structure of Fe can be dramatically altered in the outer parts of the
wind, producing a net enhancement of \FeII\ and increased strengths of the
semi-forbidden lines, implying a lower Fe abundance. The
e$^{6}$F-5p$^{6}$D transitions, arising from higher lying levels which are
formed much closer to the star's photosphere, also are available. Their
successful use as abundance diagnostics depends crucially on the
reliability of the atomic \FeII\ data. For the important infrared lines of
\FeII\ important discrepancies exist between the best two available
datasets: the Iron Project data (Seaton et al. 1994) and the data of
Kurucz (1999). First tests using an model \FeII\ atom which optimizes both
datasets for the e$^{6}$F-5p$^{6}$D transitions favor an Fe abundance not
very far from solar, a value that has been recently found by Carr et al
(1999) for the Galactic Center source IRS~7. Thus both \FeII\ line
diagnostic methods may converge to a unique iron abundance and the derived
iron abundance may be compatible with the enhanced magnesium and silicon
values.

It is necessary to perform accurate quantitative analysis to obtain more
realistic estimates of the metallicity and stellar parameters of \FMM362.  
We have begun such work following the method described in Najarro et al
(1998b); the results will be presented in detail elsewhere (Najarro et al,
in preparation).

\acknowledgements

The United Kingdom Infrared Telescope is operated by the Joint Astronomy
Centre on behalf of the U.K. Particle Physics and Research Council. We thank
the staff of the Joint Astronomy Centre for its support of these
measurements. We also are grateful to D. J. Hillier for his atmospheric
code. F. Najarro acknowledges grants of the DGYCIT under PB96-0883 and
ESP98-1351.

\clearpage

\begin{deluxetable}{ccccc}
\tablecaption{Observing Log for \FMM362. \label{tbl-1}}
\tablewidth{0pt}
\tablehead{
\colhead{UT Date} & \colhead{range} & \colhead{resolution} &
\colhead{sampling} & \colhead{Integ.} \\
\colhead{} & \colhead{(\um)} & \colhead{(\um)} & \colhead{(res.
el.)} & \colhead{(sec.)} 
} 
\startdata
19980802        & 1.90-2.54     & 0.0025  & 1/3 & 480 \\
19980802        & 1.40-2.04     & 0.0025  & 1/3 & 480 \\
19990422        & 1.87-2.51     & 0.0025  & 1/3 & 240 \\
19990422        & 3.94-4.10     & 0.00065 & 1/2 & 640 \\
19990504        & 2.10-2.18     & 0.00033 & 1/3 & 720 \\
19990504        & 1.67-1.75     & 0.00033 & 1/3 & 480 \\
\enddata
\end{deluxetable}
\clearpage

\begin{deluxetable}{ccccc}
\tablecaption{Detected Spectral Lines in \FMM362\ \label{tbl-2}}
\tablewidth{0pt}
\tablehead{
\colhead{Ion} & \colhead{Transition} & \colhead{$\lambda$(Lab)} &
\colhead{$\lambda$(Obs)} & \colhead{W$_{\lambda}$} \\
\colhead{} & \colhead{ l -- u} & \colhead{vac. \um} & \colhead{vac. \um} &
\colhead{$10^{-4}$~\um} 
}
\footnotesize
\startdata
\bf{1999 April 22:} &                                           &               &          &      \\
He I            & 3p$^{1}$P-4d$^{1}$D                           &  1.909        & 1.910    & -0.6 \\
H I             & 4 -- 8                                        &  1.945        & 1.945    &  1.4 \\
 $[$Fe II       & z$^{4}$F$_{5/2}$-c$^{4}$F$_{5/2}$             &  1.975        & 1.975    &  1.9 \\
 $[$Fe II$]$    & a$^{2}$G$_{9/2}$-a$^{2}$H$_{9/2}$             &  2.016        & 2.016    &  0.9 \\
 $[$Fe II       & z$^{4}$F$_{3/2}$-c$^{4}$F$_{3/2}$             &  2.089        & 2.089    &  2.4 \\
He I            & 3p$^{3}$P-4s$^{3}$S + 3p$^{1}$P-4s$^{1}$S     &  2.113        & 2.112    & -2.2 \\
Mg II           & 5s$^{2}$S$_{1/2}$-5p$^{2}$P$_{3/2}$           &  2.137        & 2.137    &  2.4 \\
Mg II           & 5s$^{2}$S$_{1/2}$-5p$^{2}$P$_{1/2}$           &  2.144        & 2.144    &  1.7 \\
H I             & 4 -- 7                                        &  2.166        & 2.166    &  5.2 \\
Si II           & 6p$^{2}$P$_{3/2}$-6d$^{2}$D$_{5/2,3/2}$       &  2.200 bl     & 2.201    &  0.3 \\
Na I            & 4s$^{2}$S$_{1/2}$-4p$^{2}$P$_{3/2,1/2}$       &  2.208 bl     & 2.208    &  0.6 \\
 $[$Fe II       & z$^{4}$D$_{3/2}$-c$^{4}$P$_{3/2}$             &  2.241        & 2.241    &  0.5 \\
H I             & 5 -- 27                                       &  2.360        & 2.361    &  0.6 \\
H I + Fe II     & 5 -- 26 + e$^{4}$F$_{7/2}$-5p$^{4}$D$_{5/2}$  &  2.367+2.368  & 2.368    &  0.9 \\
H I             & 5 -- 25                                       &  2.374        & 2.374    &  0.5 \\
H I             & 5 -- 24                                       &  2.383        & 2.383    &  0.6 \\
H I             & 5 -- 23                                       &  2.392        & 2.392    &  0.9 \\
H I + Mg II     & 5 -- 22 + 4d$^{2}$D$_{5/2}$-5p$^{2}$P$_{3/2}$ &  2.404+2.405  & 2.404 br &  3.1 \\
Mg II + H I     & 4d$^{2}$D$_{3/2}$-5p$^{2}$P$_{1/2}$ + 5-21    &  2.413+2.416  & 2.414 br &  2.5 \\ 
H I             & 5 -- 20                                       &  2.431        & 2.431    &  1.2 \\ 
H I             & 5 -- 19                                       &  2.449        & 2.449    &  1.0 \\ 
H I             & 6 -- 14                                       &  4.021        & 4.022    &  2.6 \\ 
He I            & 4f$^{1,3}$F-5g$^{1,3}$G                       &  4.049        & 4.050    &  5.0 \\ 
H I             & 4 -- 5                                        &  4.052        & 4.053    &  68. \\ 
                &                                               &               &          &      \\
\tablebreak
\bf{1999 May 4:} &                                              &               &          &       \\
Mg II           &  5p$^{2}$P$_{1/2}$-5d$^{2}$D$_{3/2}$          &  1.676        & 1.676    &  0.62 \\
 $[$Fe II       &   z$^{4}$F$_{9/2}$-c$^{4}$F$_{7/2}$           &  1.679        & 1.679    &  0.33 \\
Mg II           &  5p$^{2}$P$_{3/2}$-5d$^{2}$D$_{5/2}$          &  1.680        & 1.680    &  0.28 \\
H I             &   4 -- 11                                     &  1.681        & 1.681    & -0.58 \\
 $[$Fe II       &   z$^{4}$F$_{9/2}$-c$^{4}$F$_{9/2}$           &  1.688        & 1.688    &  4.0  \\
Si II           &   5s$^{2}$S$_{1/2}$-5p$^{2}$P$_{3/2}$         &  1.691        & 1.691    &  1.5  \\
Si II           &   5s$^{2}$S$_{1/2}$-5p$^{2}$P$_{1/2}$         &  1.698        & 1.698    &  0.98 \\
He I            &     3p$^{3}$P-4d$^{3}$D                       &  1.701        & 1.701    & -0.98 \\
Fe II           &     e$^{6}$F$_{11/2}$-5p$^{6}$D$_{9/2}$       &  1.704        & 1.704    &  0.12 \\
Si II           &     5f$^{2}$F-6g$^{2}$G                       &  1.719        & 1.719    &  0.23 \\
Fe II           &     e$^{6}$G$_{9/2}$-5p$^{6}$F$_{7/2}$        &  1.732        & 1.732    &  0.12 \\
Fe II           &     e$^{6}$G$_{11/2}$-5p$^{6}$F$_{9/2}$       &  1.733        & 1.733    &  0.24 \\
H I + He I      & 4 -- 10                                       &  1.737        & 1.737    & -1.2  \\
Fe II           & e$^{6}$F$_{9/2}$-5p$^{6}$D$_{9/2}$            &  1.741        & 1.741    &  0.60 \\
Mg II           &  5p$^{2}$P$_{1/2}$-6s$^{2}$S$_{1/2}$          &  1.742        & 1.742    &  1.0  \\
Mg II           &     5p$^{2}$P$_{1/2}$-6s$^{2}$S$_{1/2}$       &  1.746        & 1.746    &  1.0  \\
HeI             &   3p$^{3}$P-4s$^{3}$S                         &  2.113        & 2.113    & -0.79 \\
HeI             &   3p$^{1}$P-4s$^{1}$S                         &  2.114        & 2.114    & -0.42 \\
MgII            &   5s$^{2}$S$_{1/2}$-5p$^{2}$P$_{3/2}$         &  2.137        & 2.138    &  1.9  \\
MgII            &   5s$^{2}$S$_{1/2}$-5p$^{2}$P$_{1/2}$         &  2.144        & 2.145    &  1.0  \\
FeII            &   6s$^{6}$D$_{9/2}$-6p$^{6}$D$_{9/2}$         &  2.145        & 2.146    &  0.24 \\
HeI             &   4f$^{3}$F-7z$^{3}$Z                         &  2.165        & 2.166    & -0.19 \\
H+HeI           &   4 -- 7                                      &  2.166        & 2.167    &  4.0  \\
\enddata
\end{deluxetable}

\clearpage

\centerline{REFERENCES}

\noindent Carr, J. S., Sellgren, K. \& Balachandran, S. C. 1999, \apj, in
press

\noindent Cotera, A. S., Erickson, E. F., Colgan, S. W. J., Simpson, J.
P., Allen, D. A. \& Burton, M. G. 1996, \apj, 461, 750

\noindent Figer, D. F. 1995, Thesis, UCLA

\noindent Figer, D. F., McLean, I. S. \& Morris, M. 1996, in "The Galactic
Center", R. Gredel, ed., ASP Conf Series, 102, 263

\noindent Figer, D. F., McLean, I. S. \& Morris, M. 1999a \apj, 514, 202
(FMM99)

\noindent Figer, D. F., Morris, M., Geballe, T. R., Rich, R. M., McLean,
I. S., Serabyn, E., Puetter, R., \& Yahil, A. 1999b, \apj, 525, 759

\noindent Figer, D. F., Najarro, F., Morris, M., McLean, I. S., Geballe,
T. R., Ghez, A. M. \& Langer, N. 1998, \apj, 506, 384  (F98)

\noindent Glass, I. S., Moneti, A. \& Moorwood, A. F. M. 1990 \mnras,
242, 55P

\noindent Glass, I. S., Matsumoto, S., Carter, B. S. \& Sekiguchi, K.
1999, \mnras, 304, L10

\noindent Hillier, D. J. 1998, in "Proceedings of IAU Symposium No. 189
T. R. Bedding, A. J. Booth \& J. Davis, eds., Kluwer (Dordrecht), 209

\noindent Hillier, D. J. \& Miller, D. L. 1998, \apj, 496, 407

\noindent Langer, N., Hamann, W.-R., Lennon, M., Najarro, F., Pauldrach,
A. W. A. \& Puls, J. 1994, \aap, 290, 819

\noindent Kurucz, R., L., 1999 Kurucz CD-ROM 24, (Cambridge: Smithsonian
Astrophysical Observatory; see http://cfaku5.harvard.edu/)

\noindent Moneti, A., Glass, I. S. \& Moorwood, A. F., M. 1994, \mnras,
268, 194

\noindent Nagata, T., Hyland, A. R., Straw, S. M., Sato, S. \& Kawara, K.
1993, \apj, 406, 501 1193

\noindent Nagata, T., Woodward, C. E., Shure, M., Pipher, J. L. \& Okuda,
H. 1990, \apj, 351, 83

\noindent Najarro, F.  1995, Thesis

\noindent Najarro, F., Hillier, D. J. \& Stahl, O., 1997, \aap, 326, 1117

\noindent Najarro, F., Kudritzki, R. P., Hillier, D. J., Lamers, H. J. G.
L. M., Voors, R. H. M., Morris, P. W. \& Waters, L. B. F. M., 1998a in  
``Boulder-Munich Workshop II: Properties of hot luminous stars'', Ed. I.
D. Howarth, ASP Conf. Series, Vol 131, 57

\noindent Najarro, F., Hillier, D. J., Figer, D. F. \& Geballe, T. R.
1998b, in "The Central Parsecs, Galactic Center Workshop 1998", eds. H.
Falcke, A. Cotera, W. Huschl, F. Melia, and M. Rieke, in press.

\noindent Nota, A., Livio, M., Clampin, M., Schulte-Ladbeck, R. 1995,  
\apj, 448, 778

\noindent Okuda, H., Shibai, H., Nakagawa, T., Matsuhara, H., Kobayashi, 
Y., Kaifu, N., Nagata, T., Gatley, I. \& Geballe, T. R. 1990, \apj, 351,
89

\noindent Oliva, E.; Moorwood, A. F. M. \& Danziger, I. J. 1989, \aap, 214,
307

\noindent Seaton, M. J., Yan Y., Mihalas, D. \& Pradhan, A. K. 1994,
\mnras, 266, 805

\noindent Simon, M. Felli, M. Massi, M., Cassar, L. \& Fischer, J.  1983, 
\apj, 266, 623

\clearpage

\figcaption[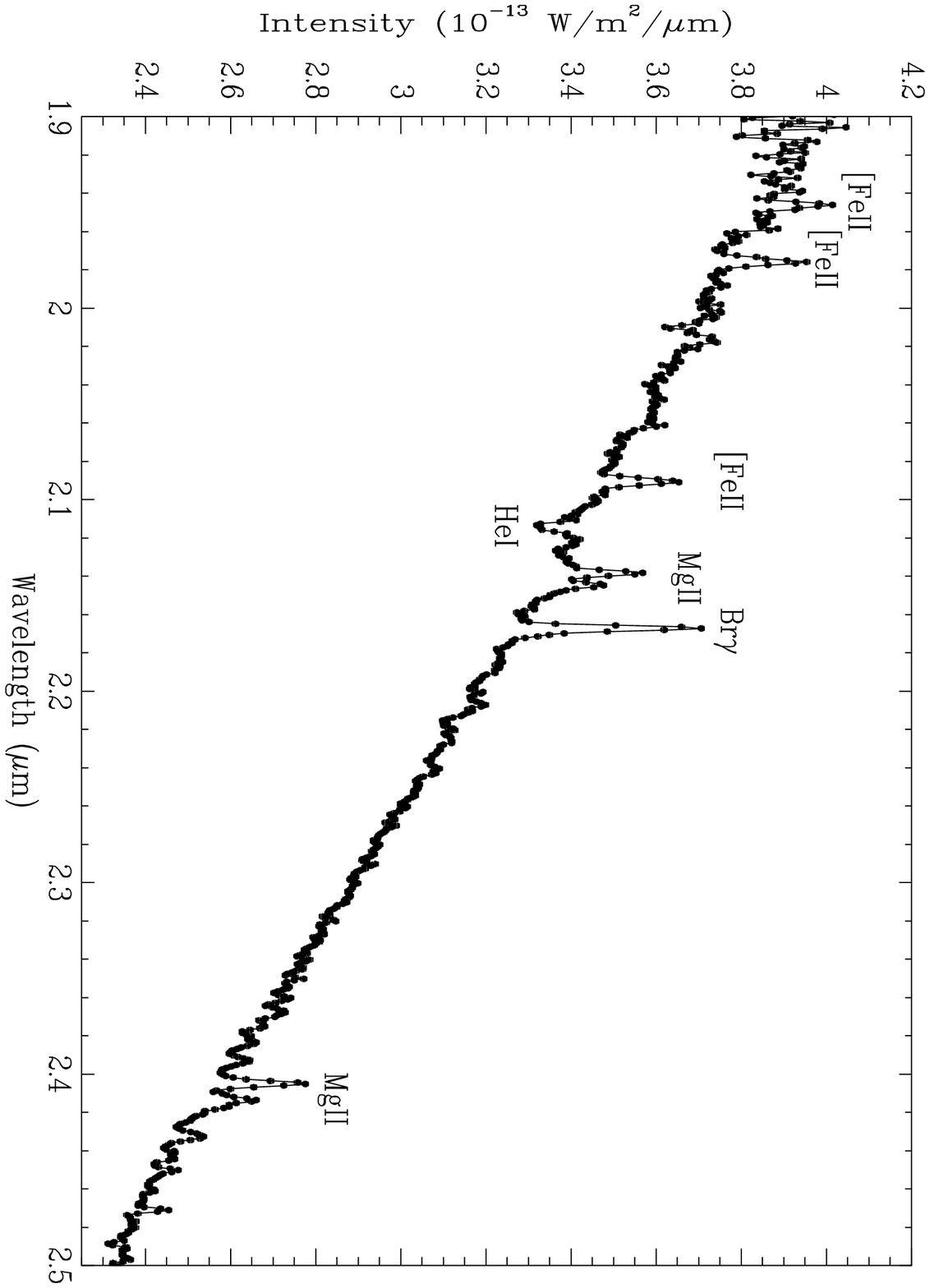]{K-band Spectrum of \FMM362, flux-calibrated using
the spectrum of the featureless Quintuplet object GCS3-2 (assumed to be a
blackbody of temperature 889~K), in the H and K bands, obtained on 1999
April 22. No dereddening was applied. The slightly smoothed spectrum has a
resolution of $\sim$0.0030~\um\ (400~\km/s). Strong lines are labelled;
see Table 2 for a complete listing of detected lines.  \label{fig1}}

\figcaption[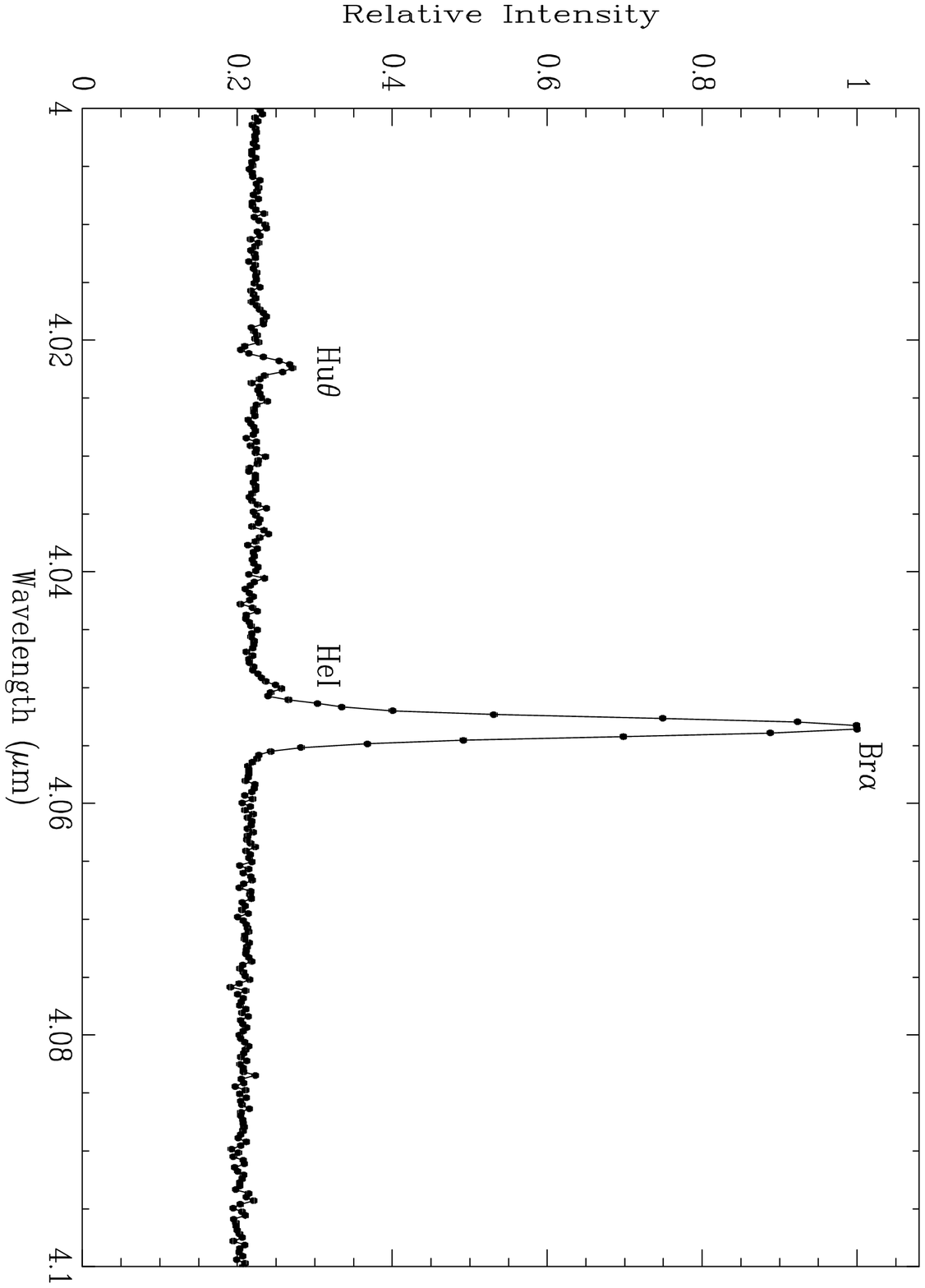]{Unsmoothed spectrum of \FMM362\ at 4~\um,
obtained on 1999 April 22, divided by that GCS3-2. The resolution is 
0.00065~\um\ (50~\km/s). \label{fig2}}

\figcaption[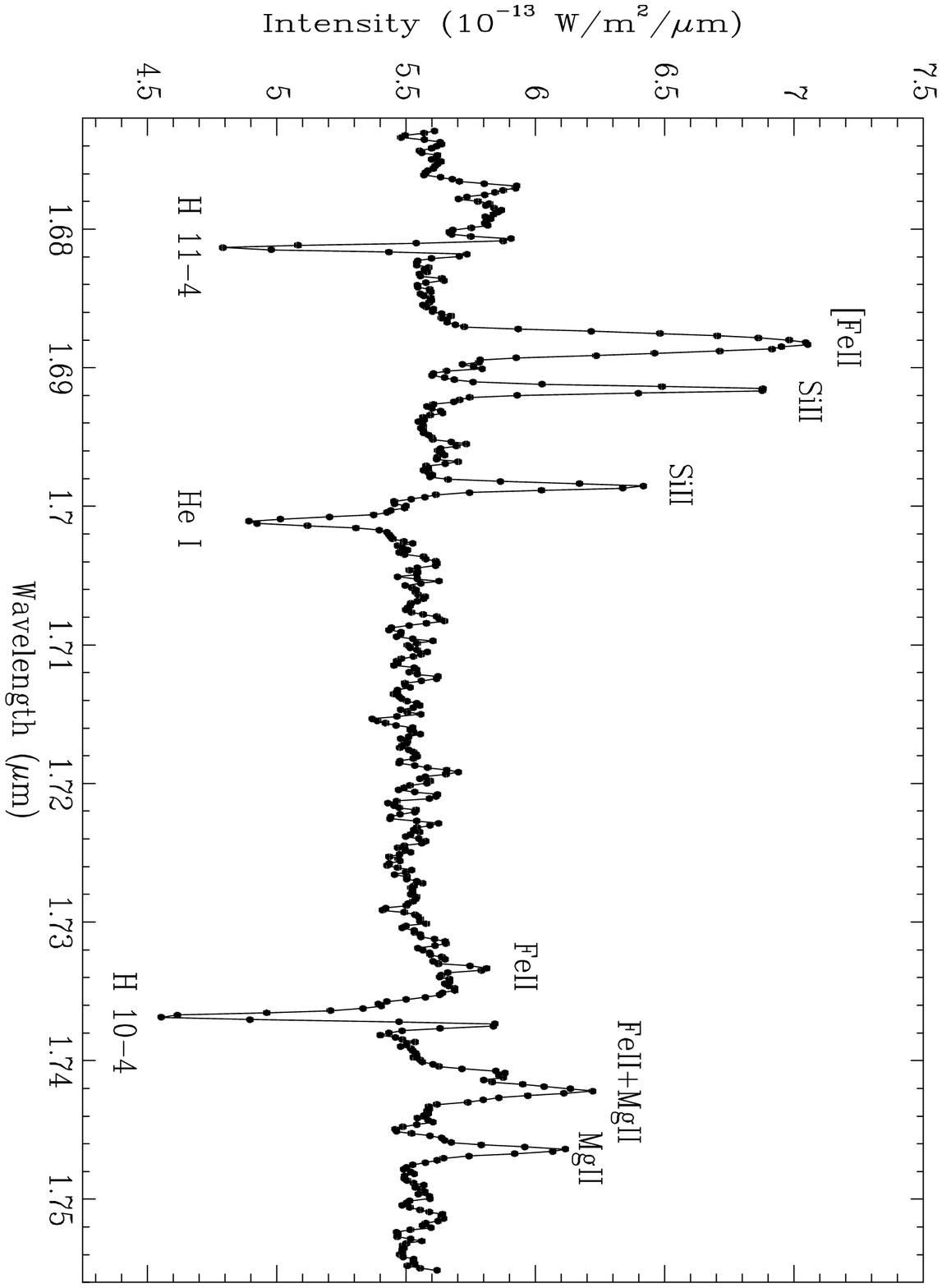]{Slightly smoothed spectrum (1.67-1.75~\um) of
\FMM362, obtained on 1999 May 4; the resolution is
0.00040~\um~($\sim$70~\km/s).  \label{fig3}}

\figcaption[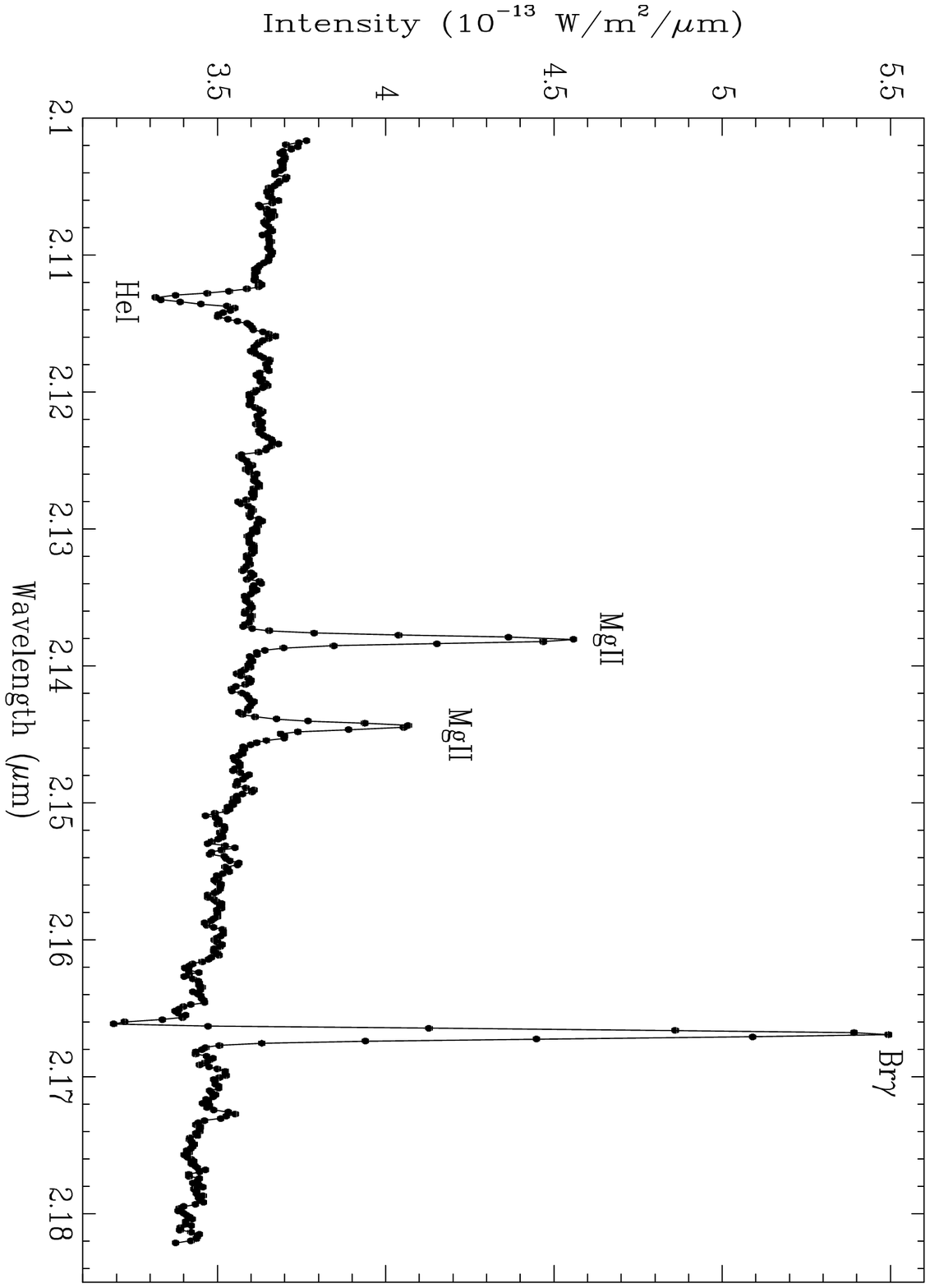]{Slightly smoothed spectrum (2.10-2.18~\um) of
\FMM362, obtained on 1999 May 4; the resolution is
0.00040~\um~($\sim$55~\km/s).  \label{fig4}}

\plotone{fig1.eps}
\plotone{fig2.eps}
\plotone{fig3.eps}
\plotone{fig4.eps}  

\end{document}